\documentclass[12pt]{article}
\usepackage{stmaryrd}
\usepackage{txfonts}
\usepackage{amsfonts}
\usepackage{amssymb}
\usepackage{latexsym}
\usepackage{graphicx}
\usepackage[english]{babel}
\input epsf.sty

\newcommand{\be}{\begin{equation}}
\newcommand{\ee}{\end{equation}}
\newcommand{\ba}{\begin{array}}
\newcommand{\ea}{\end{array}}
\newcommand{\bqa}{\begin{eqnarray}}
\newcommand{\eqa}{\end{eqnarray}}

\renewcommand{\d}{\mathrm{d}}

\def\baselinestretch{1.4}
\def \be {\begin{equation}}
\def \ee {\end{equation}}
\def \bea {\begin{eqnarray}}
\def \eea {\end{eqnarray}}
\def \nn {\nonumber}

\def \a {\alpha}
\def \b {\beta}

\def \G {\Gamma}
\def \d {\delta}

\def \m {\mu}
\def \n {\nu}
\def \k {\kappa}

\def \s {\sigma}
\def \r {\rho}
\def \o {\omega}
\def \O {\Omega}
\def \th {\theta}
\def \Th {\Theta}

\def \t {\tau}
\def \dag {\dagger}
\def \p {\partial}

\def\bd{
\begin{document}}
\def\ed{\end{document}}
\def\nn{\nonumber}
\def\bea{\begin{eqnarray}}
\def\eea{\end{eqnarray}}
\let\bm=\bibitem
\let\la=\label

\def\N{{\cal N}}
\def\sst{\scriptscriptstyle}
\def\thetabar{\bar\theta}
\def\Tr{{\rm Tr}}
\def\one{\mbox{1 \kern-.59em {\rm l}}}

%
%%%%%%%%%%%%%%%%%%%%%%%%%%%%%%%%%%%%%%%%%%%%%%%%%%%%%%%%
%%                       Abbreviations for Greek letters

\def\a{\alpha}      \def\da{{\dot\alpha}}
\def\b{\beta}       \def\db{{\dot\beta}}
\def\c{\gamma}  \def\C{\Gamma}  \def\cdt{\dot\gamma}
\def\d{\delta}  \def\D{\Delta}  \def\ddt{\dot\delta}
\def\e{\epsilon}        \def\vare{\varepsilon}
\def\f{\phi}    \def\F{\Phi}    \def\vvf{\f}
\def\h{\eta}
\def\k{\kappa}
\def\l{\lambda} \def\L{\Lambda}
\def\m{\mu} \def\n{\nu}
\def\o{\omega}
\def\P{\Pi}
\def\r{\rho}
\def\s{\sigma}  \def\S{\Sigma}
\def\t{\tau}
\def\th{\theta} \def\Th{\Theta} \def\vth{\vartheta}
\def\X{\Xeta}
\def\z{\zeta}
\def\w{\wedge}
\def\u{\underline}
\def\hs{\hspace}
\def\G{\Gamma}

%%%%%%%%%%%%%%%%%%%%%%%%%%%%%%%%%%%%%%%%%%%%
%%                      Calligraphic letters

\def\cA{{\cal A}} \def\cB{{\cal B}} \def\cC{{\cal C}}
\def\cD{{\cal D}} \def\cE{{\cal E}} \def\cF{{\cal F}}
\def\cG{{\cal G}} \def\cH{{\cal H}} \def\cI{{\cal I}}
\def\cJ{{\cal J}} \def\cK{{\cal K}} \def\cL{{\cal L}}
\def\cM{{\cal M}} \def\cN{{\cal N}} \def\cO{{\cal O}}
\def\cP{{\cal P}} \def\cQ{{\cal Q}} \def\cR{{\cal R}}
\def\cS{{\cal S}} \def\cT{{\cal T}} \def\cU{{\cal U}}
\def\cV{{\cal V}} \def\cW{{\cal W}} \def\cX{{\cal X}}
\def\cY{{\cal Y}} \def\cZ{{\cal Z}}

%%%%%%%%%%%%%%%%%%%%%%%%%%%%%%%%%%%%%%%%%%%%
%%                    Underline letters

\def\ua{\underline{\alpha}} \def\ubb{\underline{\beta}}
\def\ug{\underline{\gamma}}
\def\ub{\underline{\phantom{\alpha}}\!\!\!\beta}
\def\uc{\underline{\phantom{\alpha}}\!\!\!\gamma}
\def\um{\underline{\mu}} \def\un{\underline{\nu}}
\def\ud{\underline\delta}
\def\ue{\underline\epsilon}
\def\una{\underline a}\def\unA{\underline A}
\def\unb{\underline b}\def\unB{\underline B}
\def\unc{\underline c}\def\unC{\underline C}
\def\und{\underline d}\def\unD{\underline D}
\def\une{\underline e}\def\unE{\underline E}
\def\unf{\underline{\phantom{e}}\!\!\!\! f}\def\unF{\underline F}
\def\unm{\underline m}\def\unM{\underline M}
\def\unn{\underline n}\def\unN{\underline N}
\def\unp{\underline{\phantom{a}}\!\!\! p}
\def\unP{\underline P}
\def\unq{\underline{\phantom{a}}\!\!\! q}
\def\unQ{\underline{\phantom{A}}\!\!\!\! Q}
\def\unH{\underline{H}}
\def\ul{\underline}
%%%%%%%%%%%%%%%%%%%%%%%%%%%%%%%%%%%%%%%%%%%%
%%                  Slash letters

\def\As {{A \hspace{-6.4pt} \slash}\;}
\def\bs {{b \hspace{-6.4pt} \slash}\;}
\def\Ds {{D \hspace{-6.4pt} \slash}\;}
\def\ds {{\del \hspace{-6.4pt} \slash}\;}
\def\ss {{\s \hspace{-6.4pt} \slash}\;}
\def\ks {{ k \hspace{-6.4pt} \slash}\;}
\def\ps {{p \hspace{-6.4pt} \slash}\;}
\def\pas {{{p_1} \hspace{-6.4pt} \slash}\;}
\def\pbs {{{p_2} \hspace{-6.4pt} \slash}\;}

%%%%%%%%%%%%%%%%%%%%%%%%%%%%%%%%%%%%%%%%%%%%
%%              hatted letters

\def\Fh{\hat{F}}
\def\Vh{\hat{V}}
\def\Xh{\hat{X}}
\def\ah{\hat{a}}
\def\xh{\hat{x}}
\def\yh{\hat{y}}
\def\ph{\hat{p}}
\def\xih{\hat{\xi}}

%%%%%%%%%%%%%%%%%%%%%%%%%%%%%%%%%%%%%%%%%%%%
%%          tilde letters
\def\psit{\tilde{\psi}}
\def\Psit{\tilde{\Psi}}
\def\tht{\tilde{\th}}

\def\At{\tilde{A}}
\def\Qt{\tilde{Q}}
\def\Rt{\tilde{R}}
\def\Nt{\tilde{N}}

\def\at{\tilde{a}}
\def\st{\tilde{s}}
\def\ft{\tilde{f}}
\def\pt{\tilde{p}}
\def\qt{\tilde{q}}
\def\vt{\tilde{v}}
\def\nt{\tilde{n}}

%%%%%%%%%%%%%%%%%%%%%%%%%%%%%%%%%%%%%%%%%%%%%%%%%%%%%%%%%%
%%          bar             %%

\def\delb{\bar{\partial}}
\def\bz{\bar{z}}
\def\bD{\bar{D}}
\def\bB{\bar{B}}

%%%%%%%%%%%%%%%%%%%%%%%%%%%%%%%%%%%%%%%%%%%%%%%%%%%%%%%%%%
%%          bold                %%

\def\bk{{\bf k}}
\def\bl{{\bf l}}
\def\bp{{\bf p}}
\def\bq{{\bf q}}
\def\br{{\bf r}}
\def\bx{{\bf x}}
\def\by{{\bf y}}
\def\bR{{\bf R}}
\def\bV{{\bf V}}

%%%%%%%%%%%%%%%%%%%%%%%%%%%%%%%%%%%%%%%%%%%%%%%%%%%%%%%%%%
%%                      Miscellaneous                   %%
%%%%%%%%%%%%%%%%%%%%%%%%%%%%%%%%%%%%%%%%%%%%%%%%%%%%%%%%%%

\def\d{\delta}\def\D{\Delta}\def\ddt{\dot\delta}

\def\p{\partial} \def\del{\partial}
\def\xx{\times}
\def\uno{\mbox{1 \kern-.59em {\rm l}}}

\def\trp{^{\top}}
\def\inv{^{-1}}
\def\dag{{^{\dagger}}}
\def\pr{\prime}
\textwidth=6.0in \hoffset=-.3in \textheight=9in \voffset=-.8in
\def\baselinestretch{1.2}

\def\rar{\rightarrow}
\def\lar{\leftarrow}
\def\lrar{\leftrightarrow}

\title{Quasi-normal modes of warped black holes and warped AdS/CFT correspondence}
\author{Bin Chen and Zhi-bo Xu\footnote{Email:bchen01,xuzhibo@pku.edu.cn}\\
{\small Department of Physics} \\
{\small and State Key Laboratory of Nuclear Physics and Technology,}\\
{\small Peking University, Beijing 100871, P.R.China}}
\date{}
\bd \maketitle
\begin{abstract}
We analytically calculate the quasi-normal modes of various
perturbations of the spacelike stretched and the null warped
$AdS_3$ black holes. From AdS/CFT correspondence, these
quasi-normal modes are expected to appear as the poles in momentum
space of retarded Green's functions of dual operators in CFT at
finite temperature. We find that this is indeed the case, after
taking into account of the subtle identification of quantum
numbers. The subtlety comes from the fact that only after
appropriate coordinate transformation the asymptotic geometries of
the warped black holes are the same as the ones of warped $AdS_3$.
We show that in general the quasi-normal modes are in good
agreement with the prediction of the warped AdS/CFT
correspondence, up to a constant factor. As a byproduct, we
compute the conformal dimensions of the boundary operators dual to
the perturbations. Our results give strong support to the
conjectured warped AdS/CFT
 correspondence.
\end{abstract}

\newpage
%\begin{document}

\section{Introduction}

The study of the quasi-normal modes has been a classical subject
in black hole physics\cite{QNM}. They are called the ``sound" of
the black hole, characterizing the response of the black hole to
various perturbations. As usual, the perturbations of the black
holes obey linearized equations of motion. And the quasi-normal
modes are defined as the perturbations subject to the physical
boundary condition which states that near the horizon of the black
hole the local solution is purely ingoing and at spatial infinity
the solution is purely outgoing. As a result, the frequencies of
the perturbations are complex, indicating that the perturbations
undergo damped oscillations, just as the ring of a bell.

In fact, the frequencies of the quasi-normal modes usually take
only a discrete set of complex values. The imaginary parts of the
frequencies characterize the decay time of the perturbations. Or
in other words, their inverses characterize the relaxation time of
the system back to thermal equilibrium, with the black hole being
taken as a thermal dynamic system. Especially for the black holes
in Anti-de-Sitter(AdS) spacetime, the above picture has a nice
realization in its dual field theory. From AdS/CFT
correspondence\cite{AdSCFT}, the black hole in the bulk
corresponds to the quantum field theory on the boundary at a
finite temperature. In recent years, the correspondence at finite
temperature has been widely applied to the physical systems
ranging from quark-gluon-plasma, superconductor, superfluid to
cold atom physics\cite{Son05}. The quasi-normal modes of the black
holes correspond to the operators perturbing the thermal
equilibrium in dual field theory\cite{Witten:1998zw}. In finite
temperature field theory, the return to equilibrium under the
small perturbations is described by linear response theory. The
poles in the retarded green function of the perturbations in
momentum space encode the information of relaxation process. From
AdS/CFT correspondence, these poles are closely related to the
quasi-normal frequencies of the black holes, as first suggested in
\cite{Horowitz99}. The qualitative agreements via numerical
analysis have been found for the black holes in AdS
spacetime\cite{Horowitz99,Wang:2000gsa}. Moreover in a remarkable
paper \cite{Birmingham01}, the quantitative agreement has been
confirmed for the perturbations with various spins of the BTZ
black hole.

Actually, the Banados-Teitelboim-Zanelli(BTZ) black hole sets up
the first example of AdS/CFT correspondence. It  is a solution of
the vacuum Einstein equations in three-dimensional anti-de Sitter
spacetime\cite{BTZ}. Its dual is a two-dimensional conformal field
theory with independent left and right sectors. At thermal
equilibrium, these two sectors
 may have different temperatures $(T_L,T_R)$. For a small
 perturbation by the
 operator ${\cal O}$ with conformal weights $(h_L, h_R)$, its
 retarded Green's function has two sets of poles: \bea\label{pole}
 \o_L&=&k-4\pi iT_L(n+h_L), \nn\\
 \o_R&=&-k-4\pi iT_R(n+h_R),
 \eea
with $n$ being non-negative integer. In \cite{Birmingham01}, it
has been shown that these poles are in exact agreement with the
quasi-normal frequencies of corresponding pertubations of the BTZ
black hole.

Recently, inspired by the study of $AdS_3/CFT_2$ correspondence in
the BTZ black hole, a new kind of warped $AdS_3/CFT_2$
correspondence has been proposed in \cite{Andy08}. It was pointed
out that for the spacelike stretched and the null warped $AdS_3$
black holes, there exist dual two-dimensional conformal field
theory descriptions. The proposal was supported by the study of
thermodynamic on both sides. It was further conjectured that $v>1$
quantum topological massive gravity is holographically dual to a
two-dimensional conformal field theory with central charges
$(c_L,c_R)$. However, this correspondence is intriguing in the
sense that the warped $AdS_3$ spacetime has very different
conformal boundary from the one of $AdS_3$. The naive expectation
that the holographic CFT resides on the asymptotic boundary seems
not true any more. The dictionary in warped AdS/CFT correspondence
is not clear yet. As a step to understand the correspondence, we
calculated the scalar quasi-normal modes of the spacelike
stretched $AdS_3$ black holes in \cite{ChenXu09}. At the first
looking, the quasi-normal modes we obtained are quite different
from the prediction of the usual AdS/CFT correspondence. In this
paper, we resolve this puzzle. The key point in our analysis is to
notice that the asymptotic metric of the warped black holes is
actually not the same as the one of the warped $AdS_3$ spacetime.
One needs to make local coordinates transformations to identify
two geometries. Such transformations induce the identifications of
two sets of quantum numbers in two different backgrounds. After
using the right quantum numbers, we find that the relations on the
quasi-normal modes indeed are in good match with the prediction of
(warped) AdS/CFT correspondence.

Moreover in this paper, we  calculate other kinds of the
quasi-normal modes of the warped black holes. Our calculation
includes the vector and spinor quasi-normal modes of the spacelike
stretched warped $AdS_3$ black holes, and also the scalar, vector
and spinor quasi-normal modes of the null warped $AdS_3$ black
holes. In all these cases, the quasi-normal modes could be
obtained analytically. And after taking into account of the
subtlety on the quantum numbers, all the quasi-normal modes are in
good agreement with the prediction.

One subtle point in our study is on the boundary condition at the
asymptotical infinity. In the usual AdS black hole case, the
effective potential is infinitely high and one may impose
vanishing Dirichlet condition at infinity on the eigenfunction. In
the warped AdS black hole case, this is not obvious. The
asymptotic boundary condition on the gravitational perturbation
has been under intense
study\cite{Anninos:2009zi,Blagojevic:2009ek,Compere:2009zj} to
define the dual CFT. For other kinds of perturbations, one may
just require the flux at asymptotic boundary to be finite. This
turns out to be equivalent to the vanishing Dirichlet condition in
some cases. But generically the finite flux condition leads to
more quasi-normal modes.

To set up the correspondence, it is essential to have the
conformal dimensions of corresponding operators. One way to obtain
the conformal dimensions could be from the equations of motion of
various perturbations and studying the asymptotic behavior of the
solutions. In our case, there is another way to compute the
conformal dimension. This way stems from the existence of the
isometry algebra $SL(2,R)\times U(1)$ in the backgrounds, which
allows us to identify the conformal dimensions of primary
operators in the dual CFT.

The remaining part of the paper is organized as following: in the
next section, we focus on the spacelike stretched warped black
hole and analyze its quasi-normal modes. And in section 3, we turn
to the study of the quasi-normal modes of the null warped black
hole. In section 4, we compute the conformal dimensions of the
operators dual to the perturbations from $SL(2,R)\times U(1)$
algebra. In section 5, we end with the conclusions and the
discussions.

\section{Quasi-normal modes of the spacelike stretched warped $AdS_3$
black hole}

The spacelike stretched $AdS_3$ spacetime is the vacuum solution
of three-dimensional topological massive
gravity\cite{Deser:1981wh,Deser:1982vy}. This spacetime has an
isometry group $SL(2)_R\times U(1)_L$. It could be a stable vacuum
with appropriate boundary behavior, if the parameter $v>1$
\cite{Anninos:2009zi}. Just as the BTZ black hole could be
constructed as the orbifold of the $AdS_3$ spacetime, the black
hole asymptotic to spacelike warped $AdS_3$ could be constructed
from discrete identification as well. We would not like to review
the construction here. The interested reader can find the details
in \cite{Andy08}.

 The metric of the spacelike stretched warped $AdS_3$ black hole
takes the following form in terms of Schwarzschild coordinates:
  \bqa
    ds^2=l^2(\textrm{d}t^2+2M(r)\textrm{d}t\textrm{d}\theta+N(r)\textrm{d}\theta^2+D(r)\textrm{d}r^2),
 \eqa
where \bqa
 M(r)&=& v r-\frac{1}{2}\sqrt{r_+r_-(v^2+3)},\\
 N(r)&=&\frac{r}{4}\left(3(v^2-1)r+(v^2+3)(r_++r_-)-4v\sqrt{r_+r_-(v^2+3)}\right),\\
 D(r)&=&\frac{1}{(v^2+3)(r-r_+)(r-r_-)},
 \eqa
and $-l^{-2}$ is a negative cosmological constant and the
parameter
 $v=\mu l/3$ with $\mu$ being the mass of the graviton.
 Just like the BTZ black hole, there are two horizons located at
 $r=r_+$ and $r=r_-$. We will focus on the physical black holes without pathology,
 in which case we need to require $v > 1$. When $v=1$, there is no stretching and
 the above black hole becomes the usual BTZ black hole. This kind
 of warped black hole was discussed in \cite{Nutku:1993eb, Gurses1994}, and its
 properties were studied in \cite{Moussa:2003fc,BC07}. The other
 recent studies could be found in \cite{warped}.

 In
 \cite{Andy08}, the temperatures of the warped black holes were identified
 to be
  \be
   \frac{1}{T_H}=\frac{4\pi vl}{v^2+3}\frac{T_L+T_R}{T_R},
 \ee
 where
  \bea\label{tempwarped}
  T_L&=&\frac{(v^2+3)}{8\pi
  l}\left(r_++r_--\frac{\sqrt{(v^2+3)r_+r_-}}{v}\right), \\
  T_R&=&\frac{(v^2+3)(r_+-r_-)}{8\pi
  l},
 \eea
 are the temperature of the dual CFT. The dual two-dimensional CFT is supposed to have the central
 charges
  \be
  c_L=\frac{l}{G}\frac{4v}{v^2+3}, \hspace{5ex}
  c_R=\frac{l}{G}\frac{5v^2+3}{v(v^2+3)}.
  \ee
It has been shown in \cite{Compere:2008cv, Blagojevic:2009ek} that
the above central charges could be obtained from central extended
Virasoro algebra, based on the fact that the asymptotic symmetries
of the geometries form a semi-product of a Virasoro algebra and a
current algebra.

The scalar perturbation about this background obeys the equation
of motion: \be
  (\nabla_\mu\nabla^\mu-m^2) \Phi=0.
  \ee
  Since the background has the translational isometry along $t$
  and $\th$, we may make the following ansatz
  \be
  \Phi=e^{-i\o t+ik\th}\phi.
  \ee
After introducing the variable \be z=\frac{r-r_+}{r-r_-}, \ee we
find that the equation of motion on $\phi$ is
 \be\label{radial}
 z(1-z)\frac{d^2\phi}{dz^2}+(1-z)\frac{d\phi}{dz}+\frac{1}{(v^2+3)^2}\left(\frac{A}{z}+B+\frac{C}{1-z}\right)\phi=0,
 \ee
where
 \bea
 A&=&\frac{1}{(r_+-r_-)^2}\big(2k+\o\sqrt{r_+}(2v\sqrt{r_+}-\sqrt{v^2+3}\sqrt{r_-})\big)^2,
 \\
 B&=&-\frac{1}{(r_+-r_-)^2}\big(2k+\o\sqrt{r_-}(2v\sqrt{r_-}-\sqrt{v^2+3}\sqrt{r_+})\big)^2,\\
 C&=&3(v^2-1)\o^2-m^2l^2(v^2+3).
 \eea
The solutions take the forms of hypergeometric function. Near the
horizon, there are two  independent solutions \be
 \phi_1=z^\a (1-z)^\b F(a,b,c,z), \hspace{3ex} \phi_2=z^{-\a}
 (1-z)^\b F(a-c+1, b-c+1, 2-c, z),
 \ee
 where
 \bea
 \a&=&-i\frac{\sqrt{A}}{v^2+3}, \nn\\
 \b&=&\frac{1}{2}\left(1-\sqrt{1-\frac{4C}{(v^2+3)^2}}\right), \nn
 \eea
 and
 \bea
 c&=&2\a+1,\nn\\
 a&=&\a+\b+i\sqrt{-B}/(v^2+3),\nn\\
 b&=&\a+\b-i\sqrt{-B}/(v^2+3).\nn
 \eea

The next step is to impose the physical boundary condition to
determine the quasi-normal modes. From the definition, the
quasi-normal modes have to be purely ingoing at the horizon. The
eigenfunction $\phi_1$ satisfies this condition. On the other
hand, the asymptotical boundary condition at $z=1$ is not obvious.
In \cite{ChenXu09}, we imposed the requirement that the outgoing
flux should be finite so that the coefficients of the divergent
terms must vanish. This gives out two sets of quasi-normal modes,
determined by the relation \be
  c-a=-n, \hspace{5ex} \mbox{or}\hspace{5ex} c-b=-n,
  \ee
  with $n$ being a non-negative integer.

\begin{enumerate}
\item[1)]Case 1: $c-a=-n$\\
In this case, we are led to the following equation on $\o$:
 \bea\label{scalarcase1}
 -i\frac{1}{r_+-r_-}\frac{1}{v^2+3}(4k+\o\d)
 +\frac{1}{2}\left(1+\sqrt{1-\frac{4C}{(v^2+3)^2}}\right)=-n,
 \eea
where
 \be
 \d\equiv 2v(r_++r_-)-2\sqrt{(v^2+3)r_+r_-}.
 \ee
\item[2)] Case 2: $c-b=-n$\\
In this case, the equation on $\o$ is much simpler,
 \be\label{scalarcase2}
 -n-\frac{1}{2}+i\frac{2v\o}{v^2+3}=\frac{1}{2}\sqrt{1-\frac{4C}{(v^2+3)^2}},
 \ee
which has the solution
 \be
 \o_L=-i\left\{(2n+1)v+\sqrt{3(n+\frac{1}{2})^2(v^2-1)+(\frac{1}{4}+\frac{m^2l^2}{v^2+3})(v^2+3)}\right\}.
 \ee
Note that the frequency is pure imaginary, being independent of
the angular momentum.
\end{enumerate}

If one tries to solve $\omega$ from the equation
(\ref{scalarcase1}), one would get a solution of quite involved
form. Obviously, the quasi-normal modes look very different from
the prediction (\ref{pole}) of AdS/CFT correspondence.

However, this is just an illusion. We are going to show that the
AdS/CFT correspondence still holds, but in a subtle way. Firstly
let us consider the conformal weight of the scalar field of mass
$m$, even though the general dictionary of warped AdS/CFT
correspondence has not been set up. Consider the scalar field of
mass $m$ propagating in the spacelike warped $AdS_3$, which has
the metric \be\label{spacelikemetric}
 ds^2=\frac{l^2}{v^2+3}[-(1+r^2)d\tau^2+\frac{dr^2}{1+r^2}+\frac{4v^2}{v^2+3}
 (dx+rd\tau)^2].
 \ee
Near the boundary, with the ansatz
 \be
 \Phi=e^{i(\tilde{k}x-\tilde{\o}\tau)}\phi,
 \ee
 the scalar equation takes the form
 \be
 \p^2_z\phi+(\tilde{\o}^2-\frac{2\tilde{k}\tilde{\o}}{z}-\frac{s_s}{z^2})\phi=0,
 \ee
 where
 \be\label{s}
 s_s=\frac{3(1-v^2)}{4v^2}\tilde{k}^2+\frac{l^2}{v^2+3}m^2.
 \ee

At the asymptotical region, $\phi \sim r^{\Delta_s}$.  To have a
well-behaved solution, $\Delta_s$ should be negative. In the end,
 we have the relation
 \be\label{sweight}
 h_R=-\Delta_s=\frac{1}{2}\pm\sqrt{\frac{1}{4}+s_s}.
 \ee
This is the conformal weight of the scalar field of mass $m$.  We
will present another derivation of the conformal dimension of the
primary operator dual to the scalar perturbation in section 4.

The subtlety comes from the asymptotic behavior of the spacelike
stretched warped $AdS_3$ black hole. The metric of the asymptotic
geometry of the black hole is of the form
 \be\label{asyBH}
\frac{ds^2}{l^2}=\frac{3(v^2-1)r^2d\theta^2}{4}
+\frac{dr^2}{(v^2+3)r^2}+dt^2+2vrdtd\theta. \ee It looks different
from the spacelike warped $AdS_3$. However, after proper
identification, it is actually the same as the asymptotic geometry
of the spacelike warped $AdS_3$. Locally, the identification
is\cite{Anninos:2009zi}
 \be\label{identasy}
 \tau \leftrightarrow -\frac{v^2+3}{2}\theta, \hspace{3ex}
 x \leftrightarrow -\frac{v^2+3}{2v}t.
 \ee
This identification suggests that in the warped black hole case,
the correspondence relation should be modified. Either from the
scalar equation in (\ref{asyBH}), or from the above
identification, we have the following relations between the
quantum numbers in two backgrounds:
 \be\label{ident}
 \tilde{\o}=\frac{2}{v^2+3}k, \hspace{5ex}
 \tilde{k}=\frac{2v}{v^2+3}\o.
 \ee

 Notice that the identification
(\ref{identasy}) between the asymptotic geometries is local. In
fact, as pointed out in \cite{Anninos:2009zi} the global warped
AdS$_3$ is not the ground state of the warped black hole. In other
words, the spacelike stretched AdS$_3$ black holes with mass and
angular momentum  could not be taken as the ``excited" states.
This is reflected in the fact that the Killing vectors $\p_\t$ and
$\p_x$ of the warped AdS$_3$ spacetime are translations along the
noncompact orbits, while the Killing vector $\p_\th$ of the warped
black hole is along the compact orbit. Consequently, the quantum
number $\tilde{\o}$ is continuous, while the quantum number $k$
should be integer-valued. However, the identification
(\ref{ident}) relating these two quantities stems from the local
identification and does not care about the global properties. We
will see shortly that it is the identification (\ref{ident}) that
make the conjectured warped AdS/CFT correspondence manifest in the
context of the quasi-normal modes.

One possible resolution of the puzzle on (\ref{ident}) is that
there may exist coordinates in which the warped black hole
 has the same asymptotic geometry as the one of the warped
AdS$_3$ globally. Our belief on the existence of such coordinates
resides on the above mentioned fact that the spacelike global
stretched AdS$_3$ is not the ground state of the black holes. We
would like to take the usual point of view that the black holes
are still the ``excited" states on the vacuum, which should be
global warped spacetime in our case. If there do exist such
coordinates, then  the identification (\ref{ident}) is not
necessary any more and the warped AdS/CFT correspondence is more
transparent.

The relations (\ref{ident}) allows us to reorganize
(\ref{scalarcase1}) into
 \be\label{oR}
 \tilde{\o}_R=\frac{1}{v^2+3}(-4\pi T_Ll\tilde{k}-(i 4\pi
 T_Rl)(n+h_R)).
 \ee
This is not exactly, but quite similar to (\ref{pole}). The
discrepancy is the $1/(v^2+3)$ factor, which may be from the
warped geometry or the coordinates we choose which may induce the
redefinition of the temperature. Anyway, we would like to take
(\ref{oR}) as the convincing evidence to support the warped
AdS/CFT correspondence.

Moreover, we have another set of the quasi-normal modes determined
by (\ref{scalarcase2}). However, in this case, due to the absence
of the quantum number $k$, there is no relation on $\tilde{\o}_L$.
In fact, the relation (\ref{scalarcase2}) gives
 \be\label{k}
 \tilde{k}=-i(n+h_L),
 \ee
 where $h_L=h_R$ for scalar.
The fact that there is only one set of the quasi-normal modes
sounds strange. This could be related to the fact that the
isometry group of the spacelike warped $AdS_3$ is $U(1)_L\times SL(2)_R$. %Our result
%seems suggest that there exist only right-moving conformal field
%theory.

We would like to clarify the discrepancy between the relations
(\ref{oR},\ref{k}) and the relation (\ref{pole}) furthermore. The
strangest thing is the appearance of left-moving temperature $T_L$
in (\ref{oR}) rather than in (\ref{k}). This is mainly due to the
special property of the dual 2D CFT. In fact, the existence of
nonvanishing angular momentum in the warped black holes induce a
chemical potential in the right-moving sector. The scalar operator
in 2D CFT not only has conformal weights, but also has the right
charge coupled to the chemical potential. More precisely the
chemical potential $\O_R=-2\pi T_L, q_R=\tilde{k}$. While the
temperature $T_L$ in the relation (\ref{k}) could be recovered by
define the left-moving frequency in 2D CFT as $\o_L=2\pi T_L
\tilde{k}$. This picture is inspired by the study in Kerr/CFT
correspondence \cite{Bredberg:2009pv} and will be discussed more
clearly in our future work\cite{ChenNingXu}.

In the special case of  zero mass black hole, the left-moving
temperature $T_L \propto M^{ADT} =0$, the dual CFT becomes a
``chiral" one, similar to the one dual to the null warped
background. Now the chemical potential is zero, and the relation
(\ref{oR}) is in better match with (\ref{pole}). However, unlike
the null warped black holes we will study in the next section,
there still exist a left sector with (\ref{k}).

Another interesting point is that one could also choose the
conformal weight to be
 \be
 h_R=\frac{1}{2}-\sqrt{\frac{1}{4}+s_s},
 \ee
if $-\frac{1}{4}<s_s<0$. In this case, the finiteness of the flux
requires that
 \be
 a=-n, ~~\mbox{or}\hspace{5ex} b=-n.
 \ee
 Similarly, we get the above two relations (\ref{oR},\ref{k}). It
 is remarkable that one cannot get this conclusion from imposing the vanishing
 Dirichlet condition at asymptotic infinity. This fact suggests that the
 requirement of finite flux is not only physical but also more
 powerful.

 Before ending the discussion on the scalar quasi-normal modes, we
 would like to elucidate the organization of the quasi-normal
 relations (\ref{scalarcase1},\ref{scalarcase2}) in terms of $\tilde{\o}$ and
 $\tilde{k}$. The essential point is that the warped AdS/CFT
 correspondence states that quantum gravity asymptotic to the warped
 AdS$_3$ spacetime is holographically dual to 2D CFT. Therefore, in
 setting up the dictionary, one needs to use the quantum numbers
 of the warped AdS$_3$. More technically, the quantum number $\tilde{\o}$ and
 $\tilde{k}$ of bulk warped AdS$_3$ spacetime correspond to the
 eigenvalues of $\bar{L}_0$ and $L_0$ in dual field theory. Once we
 have the
 black hole in the bulk and change the local geometry, we should
 still use the quantum numbers $\tilde{\o}$ and
 $\tilde{k}$ in the study of warped AdS/CFT correspondence at finite temperature.

 In the remaining part of this section, we calculate the vector
 and fermionic quasi-normal modes of the spacelike AdS$_3$ black holes
 and rewrite them in terms of $\tilde{\o}$ and
 $\tilde{k}$.

\subsection{Vector perturbation}

In order to obtain the quasi-normal modes of the vector fields, we
should study the equations of the massive
 vector fields which are second order ordinal differential equations
\be
  \triangledown _{\mu}F^{\mu\nu}=m^2A^{\nu}.
  \ee
  However, one may work with the following first order equations whose solutions
  are the solutions of the above equations in three-dimensional
  spacetime:
  \be
   \epsilon_{\lambda}^{\
   \alpha\beta}\partial_{\alpha}A_{\beta}=-mA_{\lambda},\label{vector}
  \ee where $\epsilon_{\lambda}^{\ \alpha\beta}$ is the Levi-Civita
tensor with $\epsilon^{tr\theta}=1/\sqrt{-g}$. Since the
background have translational symmetries along $t$ and $\theta$,
we can make the following ansatz \be
    A_{\mu}=e^{-i\omega t+ik\theta}\phi_{\mu}.
\ee
Then the equations of motion can be given explicitly
 \bqa
 \frac{\textrm{d}\phi_t}{\textrm{d}r}&=&2D(r)\left((\frac{-\omega k}{m l}+m l M(r))\phi_t-(\frac{\omega^2}{m
 l}+m l)\phi_\theta\right),\label{eom1}\\
 \frac{\textrm{d}\phi_\theta}{\textrm{d}r}&=&2D(r)\left ((\frac{k^2}{m
 l}+m l N(r))\phi_t-(\frac{-\omega k}{m
 l}+m l M(r))\phi_\theta\right),\\
 \phi_r&=&-\frac{2D(r)}{m l}(ik\phi_t+i\omega\phi_\theta). \eqa
 After changing the variables to
 \be
  z=\frac{r-r_+}{r-r_-},
\ee we have the second order ordinary equation for $\phi_t$
 \be
 z(1-z)\frac{d^2\phi_t}{dz^2}+(1-z)\frac{d\phi_t}{dz}+(\frac{A_v}{z}+B_v+\frac{C_v}{1-z})\phi_t=0,
\ee where \bqa
A_v&=&\frac{1}{(r_+-r_-)^2(v^2+3)^2}\left(2k+\omega\sqrt{r_+}(2v\sqrt{r_+}-\sqrt{v^2+3}\sqrt{r_-})\right)^2,\nn
 \\
 B_v&=&-\frac{1}{(r_+-r_-)^2(v^2+3)^2}\left(2k+\omega\sqrt{r_-}(2v\sqrt{r_-}-\sqrt{v^2+3}\sqrt{r_+})\right)^2,\nn\\
 C_v&=&\frac{1}{(v^2+3)^2}\left(3(v^2-1)\omega^2-(m^2l^2+2mv
 l)(v^2+3)\right).
\eqa The solutions can be written in terms of hypergeometric
functions. There are two independent solutions,
 \be
 \phi_1=z^{\alpha_v} (1-z)^{\beta_v+1} F(a_v+1,b_v+1,c_v,z), \hspace{3ex}
 \phi_2=z^{-\alpha_v}
 (1-z)^{\beta_v+1 }F(a_v-c_v+2, b_v-c_v+2, 2-c_v, z),
 \ee
where $\alpha_v=-i\sqrt{A_v},\  \beta_v=(-1+\sqrt{1-4C_v})/2$ and
 \be c_v=1+2\alpha_v,\ \ a_v=\alpha_v+\beta_v+i\sqrt{-B_v},\ \
 b_v=\alpha_v+\beta_v-i\sqrt{-B_v}.
 \ee
 Since the quasi-normal modes are purely ingoing at the horizon,
 $\phi_1$ is the solution we need.

 %However the second ordinary
 %equation for $\phi_\theta$ is much involved, it is difficult to solve the equation directly.
 %Fortunately we can obtain the
 %solution in terms of $\phi_t$
 By using the equation (\ref{eom1}), we obtain $\phi_\theta$
 % which can be written as follow
 in terms of the variable $z$ ,
\be \phi_\theta=\tilde A_v\phi_t+\tilde B_v
\frac{1}{1-z}\phi_t+\tilde C_v
z\frac{\textrm{d}\phi_t}{\textrm{d}z},
 \ee
 where
 \bqa
 \tilde A_v&=&\frac{1}{2\omega^2+2m^2l^2}(-2\omega k+2m^2l^2vr_--m^2l^2\sqrt{r_+r_-(v^2+3)}),\nn\\
 \tilde B_v&=&\frac{m^2l^2v(r_+-r_-)}{\omega^2+m^2l^2},\nn\\
 \tilde C_v&=&-\frac{m l(v^2+3)(r_+-r_-)}{2\omega^2+2m^2l^2}.\nn
 \eqa
Similarly the solution can be written in terms of  hypergeometric
functions explicitly. Finally we have
 \bqa
  \phi_\theta&=&z^{\alpha_v}(1-z)^{\beta_v}\left\{(\tilde A_v+\tilde
  C_v(\alpha_v+\beta_v-b_v))(c_v-b_v-1)F(a_v,b_v,c_v,z)\right.\nonumber\\
  & &+\left(2\beta_v(\tilde A_v+\tilde C_v(\alpha_v+\beta_v-b_v))+a_v\tilde
  C_v
  (c_v-a_v-1)\right)F(a_v,b_v+1,c_v,z)\nonumber\\
 & &\left. +a_v\frac{m l(r_+-r_-)}{2\omega^2+2m^2l^2}\left(2m l
 v-(v^2+3)\beta_v\right)F(a_v+1,b_v+1,c_v,z)\right\}\label{phith}\\
  \phi_t&=&a_vz^{\alpha_v} (1-z)^{\beta_v+1} F(a_v+1,b_v+1,c_v,z).
 \eqa

   There are two special cases we would like to consider separately. If $a_v=0$ or $c_v-b_v-1=0$
   which may lead to $\o^2+m^2l^2=0$, one can directly solve the equations.
   Here we introduce a new parameter $\tilde\b_v=\frac{2mlv}{v^2+3}$
   for convenience.
   In the case $\o=-iml$, we have $\tilde\b_v=\b_v$ which means $a_v=0$.
   In this case, one of the
   solutions of the equations of motion is
   \bea\label{av}
     \phi_t=0, \ \ \ \phi_\th=z^{\alpha_v}(1-z)^{\tilde\beta_v}.
   \eea
This solution has purely ingoing mode at the horizon and so is the
right solution we want. While the other solution is that $
\phi_t=z^{-\alpha_v}(1-z)^{-2mlv}$ and
   $\phi_\theta$ can be obtained from the equations of motion correspondingly.
   One can find that  $\phi_\th$ approaches to
   $z^{-\alpha_v}$ near the horizon and $(1-z)^{-1-2mlv}$ at
   infinity. This is a solution with outgoing mode near the
   horizon. Thus it is not the solution we need.

   In the case $\o=iml$,  one has $c_v-b_v-1=0$.
    Then one
   solution is $\phi_t=0, \ \ \
   \phi_\th=z^{-\alpha_v}(1-z)^{\b_v}$. It is a solution with
   outgoing mode. The other solution is a solution with ingoing
   mode, where $\phi_t=z^{\alpha_v}(1-z)^{-2mlv}$ and $\phi_\th$
   approaches to $(1-z)^{-1-2mlv}$ at asymptotic infinity.

One has to impose the physical boundary condition at asymptotic
infinity. One may require the flux vanishing condition as the
scalar field case. Let us consider the energy flux and the angular
momentum flux. They are defined as
  \be
    \mathcal {F}_e=\int dtd\th\sqrt{-g}T^r_t,\ \
    \mathcal {F}_a=\int dtd\th\sqrt{-g}T^r_\th.
  \ee
  Using the equations of motion and considering the flux at
  infinity after time averaging, we have
  \bqa
     \mathcal {F}_e\simeq (ik\phi_t+i\o\phi_\th)\phi_t^*+c.c.,\\
     \mathcal {F}_a\simeq (ik\phi_t+i\o\phi_\th)\phi_\th^*+c.c..
  \eqa

   For $\o$ is real, the leading term of the flux at asymptotic infinity is proportional to
   \be
      \left|\frac{\Gamma(c_v)\Gamma(a_v+b_v+2-c_v)}{\Gamma(a_v+1)\Gamma(b_v+1)}\right|^2(1-z)^{-1-2\b_v}
   \ee
  Generally the finite flux
boundary condition for the solutions at asymptotic infinity gives
the following relation:
  \be\label{vectorcase}
     b_v+1=-n,~~ \textrm{or}\ \hspace{3ex} a_v=-n.
  \ee
Note that $a_v=0$ also satisfies the boundary condition from the
solution (\ref{av}).

For $v=1$, we need some special considerations since the third
line of (\ref{phith}) may be zero. In the end we find that it
leads to the same quasi-normal modes as the ones in the BTZ black
hole, once the proper redefinition of the temperature is taken
into account.

\begin{enumerate}
\item For the case $b_v+1=-n$, we have \bea\label{vectorcase1}
 -i\frac{1}{r_+-r_-}\frac{1}{v^2+3}(4k+\o\d)
 +\frac{1}{2}\left(1+\sqrt{1-4C_v}\right)=-n,
 \eea
 which is very similar to (\ref{scalarcase1}) except replacing $C$
 with $C_v(v^2+3)^2$.

 \item For the case $a_v=-n$, we have
\be\label{vectorcase2}
 i\frac{2v\o}{v^2+3}=n-\frac{1}{2}+\frac{1}{2}\sqrt{1-4C_v},
 \ee
 \end{enumerate}

For the general case $v>1$, the relations
(\ref{vectorcase1},\ref{vectorcase2}) lead to quite involved forms
of the frequencies of the quasi-normal modes. However, as we
stated above, we have to consider the subtle identification of the
quantum numbers. As the first step to compare the above results
with the CFT prediction, we would like to discuss the conformal
weights of the massive vectors. From the massive vector field
equation in the spacelike stretched $AdS_3$ spacetime, after
analyzing the behavior of the solution at asymptotic infinity, we
can determine its conformal weight. One choice is
 \be\label{vweight}
 h_R^v=\frac{1}{2}+\sqrt{\frac{1}{4}+s_v}
 \ee
 with
 \be
 s_v=\frac{3(1-v^2)}{4v^2}\tilde{k}^2+\frac{(m^2l^2+2vml)}{v^2+3}.
 \ee

Therefore, taken into account of the subtle identification
(\ref{ident}), the above relation (\ref{vectorcase}) could be
rewritten as
 \bea\label{oRk}
 \tilde{\o}_R^v&=&\frac{1}{v^2+3}(-4\pi T_Ll\tilde{k}-(i 4\pi
 T_Rl)(n+h_R^v)),~~\mbox{or}\\
 \tilde{k}&=&-i(n+h_L^v).
 \eea
where $h_L^v=h^v_R-1$.

 As the scalar case, there is another choice of the conformal
weight
 \be\label{hRv}
 h_R^v=\frac{1}{2}+\sqrt{\frac{1}{4}+s^\prime_v},
 \ee
 where
 \be
s_v^\prime=\frac{3(1-v^2)}{4v^2}\tilde{k}^2+\frac{(m^2l^2-2vml)}{v^2+3},
 \ee
 which corresponds to the vector field with a different helicity.
 In this case, the vanishing boundary
 condition cannot give the right constraint. One has to read from
 flux finiteness condition. In the end, one obtains the same
 relations (\ref{oRk}) with $h_L^v=h^v_R+1$.

\subsection{Fermion perturbation}

  In this subsection we analyze the quasi-normal modes of the fermionic
  fields on the spacelike stretched warped AdS$_3$ black holes
background. In order to solve the Dirac equations, we should
choose the vielbein for the background spacetime and calculate the
corresponding spin connection.

The vielbein $e^a_\mu$ is chosen as  \bqa
  e^0=\frac{l}{2\sqrt{D(r)}}\textrm{d}\theta,\ \ e^1=l\sqrt{D(r)}\textrm{d}r,\ \
  e^2=ldt+M(r)l\textrm{d}\theta,
\eqa where $e^a_{\mu}\textrm{d}x^{\mu}=e^a$. The spin connection
can be calculated straightforwardly. The nonvanishing components
of the spin connection are
 \bqa
   \omega^{01}_t=-\omega^{10}_t=-M^{\prime},\ \ \
   \omega^{02}_r=-\omega^{20}_r=-\sqrt{D}M^{\prime},\nn\\
   \omega^{01}_{\theta}=-\omega^{10}_{\theta}=MM^{\prime}-N^{\prime},\
   \ \
   \omega^{12}_{\theta}=-\omega^{21}_{\theta}=\frac{-M^{\prime}}{2\sqrt{D}}.\nn
\eqa

 The Dirac equations are
 \bqa  \gamma^a
e_a^{\mu}(\partial_{\mu}+\frac{1}{2}\omega_{\mu}^{ab}\Sigma_{ab})\Psi+m\Psi=0,
\eqa
 where
$\Sigma_{ab}=\frac{1}{4}[\gamma_a,\gamma_b],\
\gamma^0=i\sigmaup^2,\ \gamma^1=\sigmaup^1,\ \gamma^2=\sigmaup^3.$
Similarly, we change to the variable  \bqa
 z=\frac{r-r_+}{r-r_-},\nn
\eqa and with the ansatz $\Psi=(\psi_+, \psi_- )e^{-i\omega
t+ik\theta}$, we then have   \bqa
  \frac{\textrm{d}\psi_{\pm}}{\textrm{d}z}&=&\left(\pm\frac{2i(\omega vr_+ -\omega e+k)}{z(v^2+3)(r_+-r_-)}
    \pm\frac{2i\omega v}{(v^2+3)(1-z)}-\frac{1}{2(1-z)}-\frac{1}{4z}\right)\psi_{\pm}\nn\\
  &&+(\mp
  i\omega+ml-\frac{v}{2})\frac{1}{\sqrt{(v^2+3)z}(1-z)}\psi_{\mp},
\eqa where $e=\frac{1}{2}\sqrt{r_+r_-(v^2+3)}$.

The solutions of these equations with only ingoing flux at the
horizon are given by hypergeometric functions
 \bqa
  \psi_+&=&z^{\alpha_f+1/2}(1-z)^{\beta_f}F(a_f+1,b_f,c_f+1,z),\\
   \psi_-&=&\frac{c_f(i\omega+ml-v/2)}{a_f(b_f-c_f)\sqrt{v^2+3}}z^{\alpha_f}(1-z)^{\beta_f}F(a_f,b_f,c_f,z),
\eqa where $c_f=2\alpha_f+1,\ a_f=\alpha_f+\beta_f+\gamma_f,\
b_f=\alpha_f+\beta_f-\gamma_f$, and
 \bqa
 \alpha_f&=&-\frac{2i(\omega v r_+ -\omega
 e+k)}{(v^2+3)(r_+-r_-)}-\frac{1}{4},\\
 \beta_f&=&\frac{1}{2}-\sqrt{\frac{(m
 l-v/2)^2}{v^2+3}-\frac{3\omega^2(v^2-1)}{(v^2+3)^2}},\\
 \gamma_f&=&\frac{2i(\omega v r_- -\omega
 e+k)}{(v^2+3)(r_+-r_-)}-\frac{1}{4}.
 \eqa

 Similar to the vector case, there are two special cases: $c_f-b_f=0$ and $a_f=0$, both of which  lead to
   $\o^2+(ml-\frac{1}{2})^2=0$. In fact, if $Im\o\leq 0$ then one has
   $c_f-b_f=0$, while we have $a_f=0$ for $Im\o \geq 0$.

    For the case $i\o=ml-\frac{1}{2}$, one
   has a solution with purely ingoing mode near the horizon as $\psi_+=0,\
   \psi_-=z^{\alpha_f}(1-z)^{\tilde\beta_f}$, where
   $\tilde\b_f=\frac{(2ml-v)v}{v^2+3}+\frac{1}{2}$. The other solution
   has outgoing mode near the horizon.

    For the case $-i\o=ml-\frac{1}{2}$, one solution with outgoing mode
    is $\psi_-=0, \ \psi_+=z^{-\a_f-\frac{1}{2}}(1-z)^{\tilde\b_f}$.
    The other one which has ingoing mode is $\psi_+=z^{\a_f}(1-z)^{1-\tilde\b_f}$
    and $\psi_-$ has the same asymptotic behavior of $\psi_+$ near
    the horizon and at infinity.

 We impose the vanishing flux condition at asymptotic infinity. The flux is
 \bqa
 \sqrt{-g}\bar\Psi e^r_1\gamma^1\Psi\cong
 (1-z)^{-1}(|\psi_+|^2-|\psi_-|^2).
 \eqa
 The leading divergent term of the flux is of order
 $(1-z)^{2\beta_f-1}$, where if $\beta_f$ is complex we choose $Re(\beta_f)<1/2$
 for the branch cut. Its coefficient
 \bqa
  \left|
  \frac{\Gamma(c_f+1)\Gamma(c_f-a_f-b_f)}{\Gamma(c_f-a_f)\Gamma(c_f-b_f+1)}\right|^2
 \eqa
 must vanish. Considering that  the solution in the $c_f-b_f=0$ case  also satisfy
 the boundary condition,
  so we have
 \bqa
   c_f-a_f=-n,\ \mbox{or}\ \ c_f-b_f=-n.
 \eqa
 for the quasi-normal modes.
\begin{enumerate}
\item In the case: $c_f-b_f=-n$, we have
 \be\label{spinor1}
 -n+i\frac{2v\omega}{v^2+3}=\sqrt{\frac{(m
 l-v/2)^2}{v^2+3}-\frac{3\omega^2(v^2-1)}{(v^2+3)^2}},
 \ee

\item In the case: $c_f-a_f=-n$, we have \bqa\label{spinor2}
 -i\frac{l}{r_+-r_-}\frac{1}{v^2+3}(4k+\omega\delta)
 +\frac{1}{2}+\sqrt{\frac{(m
 l-v/2)^2}{v^2+3}-\frac{3\omega^2(v^2-1)}{(v^2+3)^2}}=-n.
 \eqa
 \end{enumerate}
%where
% \be
% \delta=2v(r_++r_-)-2\sqrt{(v^2+3)r_+r_-}.
% \ee

In the limit $v=1$, the spectrum of the quasi-normal mode is
 \bqa
 \omega_R&=&-\frac{4k}{\delta}-i\frac{4(r_+-r_-)}{\delta}(n+\frac{1}{2}(1+|m
 l-1/2|)),\\
 \omega_L&=&-2i(n+\frac{1}{2}|m l-1/2|)
 \eqa
where $\delta=2(\sqrt{r_+}-\sqrt{r_-})^2$. So the left and the right
conformal weights are given by $h_L=|ml-1/2|/2$, and
$h_R=(1+|ml-1/2|)/2$. This is in precise match with the results in
the BTZ black hole obtained in \cite{Birmingham01}, after
considering
the different choice of the vielbeins.% which has a discrete space
%inverse transformation except for other continued transformations.

For the general case with $v>1$, in order to compare with the
prediction of warped AdS/CFT correspondence, we need the conformal
weights of the massive fermionic operators. In the similar spirit
as the scalar and the vector, we have
 \be
 h_R^f=\frac{1}{2}+\sqrt{\frac{(m
 l-v/2)^2}{v^2+3}-\frac{3\tilde{k}^2(v^2-1)}{(v^2+3)^2}}.
 \ee
Taken into account of the identification (\ref{ident}), the
relation (\ref{spinor2}) is of the form \be\label{f1}
\tilde{\o}^f_R=\frac{1}{v^2+3}(-4\pi T_Ll\tilde{k}-(i 4\pi
 T_Rl)(n+h_R^f)), \ee and the relation
(\ref{spinor1}) is of the form \be\label{f2}
 \tilde{k}^f=-i(n+h_L^f),
 \ee
where $h^f_L=h^f_R-\frac{1}{2}$.

Similarly, one may have \be
 h_R^f=\frac{1}{2}+\sqrt{\frac{(m
 l+v/2)^2}{v^2+3}-\frac{3\tilde{k}^2(v^2-1)}{(v^2+3)^2}},
 \ee
 and $h^f_L=h^f_R-\frac{1}{2}$. Even in this case, the
 quasi-normal modes are still given by (\ref{f1},\ref{f2}).

Let us summarize the result we obtained in this section. No matter
what kind of the perturbations we considered, the quasi-normal
modes of the spacelike stretched $AdS_3$ black hole could be
simply written as
 \bea
 \tilde{\o}_R&=&\frac{1}{v^2+3}(-4\pi T_Ll\tilde{k}-(i 4\pi
 T_Rl)(n_1+h_R)),\\
 \tilde{k}&=&-i(n_2+h_L),
 \eea
with $n_1,n_2$ being non-negative integers.

\section{Quasi-normal modes of the null warped black holes}

Null warped $AdS_3$ spacetime is another vacuum solution of
three-dimensional topological massive gravity\footnote{For the
related study on AdS wave solution, see \cite{AyonBeato:2004fq}.}.
It is only well defined at $v=1$. Similar to other warped $AdS_3$
spacetime, it also has isometry group $SL(2,R)\times U(1)_{null}$.
The null warped black hole could be taken as the quotient of the
null warped $AdS_3$. The  metric of the null warped black hole is
of the form
 \be
   \frac{ds^2}{l^2}=-2r\textrm{d}\theta\textrm{d} t+(r^2+r+\alpha^2)\textrm{d}
   \theta^2+\frac{\textrm{d} r^2}{4r^2},
 \ee
 where $1/2>\alpha>0$ in order to avoid the naked causal
 singularity. The horizon is located at $r=0$.
 From the thermodynamics of this black hole, it was argued that
 there exist non-vanishing right-moving temperature
 \be
 T_R=\frac{\a}{\pi l}.
 \ee

One may propose the following conjecture:  $v=1$ quantum
topological massive gravity with  asymptotical null warped $AdS_3$
geometry is holographically dual to a 2D boundary CFT with the
left-moving central charge $c_L=\frac{l}{G}\frac{4v}{v^2+3}$ and
the right-moving central charge $c_R=\frac{(5v^2+3)l}{Gv(v^2+3)}$.
From the black hole entropy, it seems that it is not necessary to
have left-moving central charge since $T_L=0$. However, the
diffeomorphism anomaly requiring that $c_L-c_R=-\frac{l}{Gv}$ asks
for the existence of the left-moving sector.

 In order to check this conjectured correspondence, we study the
quasi-normal modes in the null warped black hole in this section.
Firstly let us consider the scalar perturbation.  The equation of
motion for the scalar field is
 \be
   \nabla^2\Phi-m^2\Phi=0.
\ee
  Taken the ansatz $\Phi=e^{-i\omegaup t+ik\theta}R(r)$, the
  equation
 becomes
 \be
   \frac{\textrm{d} }{\textrm{d} r}(4r^2\frac{\textrm{d}}{\textrm{d} r}R)+(-\frac{2\omegaup
   k}{r}+\frac{\omegaup^2(r^2+r+\alpha^2)}{r^2}-m^2l^2)R=0.
 \ee
 The above equation can be solved by Kummer
 functions
 \be
   R_{\pm}=e^{-\frac{z}{2}}z^{\frac{1}{2}\pm\tilde m_s}F(\frac{1}{2}\pm\tilde m_s-\k,1\pm2\tilde m_s, z)
 \ee
 where $z=-i\omegaup\alpha\frac{1}{r},\ \tilde
 m_s=\frac{1}{2}\sqrt{1+m^2l^2-\omegaup^2}$ and $
 \k=\frac{i}{4\alpha}(\omegaup-2k)$. Here we choose
 $-\pi<argz<\pi$ for the branch cut. Actually the solution should be
 a combination
 \be\label{combined}
    R=C_1R_++C_2R_-.
 \ee

 Now let us consider the boundary condition for the quasi-normal modes. There
 have to be only ingoing modes near the horizon where $z$ approaches to
 the infinity. The asymptotic expansion of Kummer function at asymptotic infinity is
 \be
   F(\alpha,\gamma,z)\sim \
   \frac{\Gamma(\gamma)}{\Gamma(\gamma-\alpha)}e^{-
   i\alpha\piup}z^{-\alpha}+\frac{\Gamma(\gamma)}{\Gamma(\alpha)}e^z
   z^{\alpha-\gamma}.
 \ee
 where it requires $-\frac{3\pi}{2}<argz<\frac{\pi}{2}$. In our case, the second term of the right hand equation
 corresponds to the outgoing modes near the horizon. For the
 solution (\ref{combined}), the vanishing outgoing condition
 requires
  \be
    C_1=-\frac{\G(1-2\tilde m_s)}{\G(\frac{1}{2}-\tilde m_s-\k)}C,
    \ \ C_2=\frac{\G(1+2\tilde m_s)}{\G(\frac{1}{2}+\tilde
    m_s-\k)}C,
  \ee
  where $C$ is a constant.
  Next we require the flux at infinity to be vanishing. The flux is given by
  \be
    \mathcal {F}\sim \frac{2\pi}{i}r^2(\Phi^*\p_r\Phi-c.c.)
  \ee
   For the solution $R(z)$, the leading term of the corresponding flux
   is proportional to $C_1^*C_2-c.c.$ and the sub-leading term is
   proportional to
   $C_2C_2^*r^{-1+2\tilde m_s}$. So if $Re(\tilde m_s)>\frac{1}{2}$,
   the flux vanishes if $C_2=0$ that is
 \be
   \frac{1}{2}+\tilde m_s-\k=-n\label{squasi}
 \ee
   We will see that $\frac{1}{2}+\tilde m_s$ is the conformal weight
   of the scalar of mass $m$ in the following and next section.
  While for the case $Re(\tilde
  m_s)<\frac{1}{2}$,  the flux vanishes when $C_1=0$ or $C_2=0$, which
  indicate that
\be
   \frac{1}{2}+\tilde m_s-\k=-n, \textrm{or} \ \ \frac{1}{2}-\tilde
   m_s-\k=-n.
 \ee
 Note that for the case $Re(\tilde
  m_s)<\frac{1}{2}$, there are two possible choices of the
  conformal weights after considering the identification of
  the quantum numbers: $h_R=\frac{1}{2}\pm\tilde m_s$. Therefore,
  the above relations on quasi-normal frequencies could be simply
  written into
 \be
 h_R-\k =-n.
 \ee

%   So it may have two sets of quasi normal modes corresponding two
%   different conformal weight.
   %For $r$ approaching to infinity, the asymptotic flux of the scalar field
% must vanish. The leading term of field at asymptotic infinity is of order $z^{\frac{1}{2}+\tilde
 %m_s}$. In order to have well behaved property, we need $\frac{1}{2}+\tilde
 %m_s>0$. In fact, this quantity is just the conformal weight $h_R$.
 %If we only consider the lower frequency quasi-normal modes that
 %satisfy
 % $m^2l^2>|\omegaup^2|$, we must choose
 % $\tilde m_s=\frac{1}{2}\sqrt{1+m^2l^2-\omegaup^2}$.

  In order to compare the quasi-normal mods with the result from the dual conformal field
  theory, we first need to identify the conformal weights of the
  dual operators. Now we calculate the conformal weight of the scalar field of
  mass $m$ from its asymptotic behaviors. The metric of the null warped $AdS_3$ spacetime could be of the following form
 \be\label{nullmetric}
 \frac{ds^2}{l^2}=\frac{du^2}{u^2}+\frac{dx^+dx^-}{u^2}+\left(\frac{dx^-}{u^2}\right)^2.
 \ee
  Near the asymptotic region, we have the scalar equation of motion:
  \be\label{nullscalar}
  4r^2\frac{d^2}{dr^2}R+8r\frac{d}{dr}R+(4\tilde{k}_n^2-m^2l^2)R=0,
  \ee
where we have introduced $u^2=1/r$ and have made the following
ansatz: \be \Phi=e^{-i\tilde{\o}_nx^-+i\tilde{k}_nx^+}R(r). \ee
The subscript $n$ is introduced to denote the null background.
 The eigenfunction is of
the form $r^{\Delta_s}$ with $-\Delta_s$ being the conformal
weight. The equation (\ref{nullscalar}) leads to
 \be
 \Delta_s^2+\Delta_s + \frac{4\tilde{k}_n^2-m^2l^2}{4}=0,
 \ee
 with the solution
 \be
 \Delta_s=-\frac{1}{2} \pm\frac{1}{2} \sqrt{1+m^2l^2-4\tilde{k}_n^2}.
 \ee
 In order that the solution is well-behaved, $\Delta$ should be
 negative, this helps us to take
 \bea
 \Delta_s=-\frac{1}{2} \pm\frac{1}{2} \sqrt{1+m^2l^2-4\tilde{k}_n^2}, \hspace{5ex}\mbox{if
 $m^2<4\tilde{k}_n^2$,}\nn\\
 \Delta_s=-\frac{1}{2}-\frac{1}{2} \sqrt{1+m^2l^2-4\tilde{k}_n^2}, \hspace{5ex}\mbox{if
 $m^2>
 4\tilde{k}_n^2$.}
 \eea
The conformal dimension of the dual primary operator is just
$h_R^s=-\Delta_s$.

Similar to the spacelike warped case, in order to compare with the
prediction of dual CFT, one has to consider the identification of
quantum numbersdue to the coordinates transformations. In the
asymptotic region, we can make the following identification
locally:
 \be
 u^2\leftrightarrow \frac{1}{r}, \hspace{3ex}x^-
 \leftrightarrow \theta, \hspace{3ex} x^+\leftrightarrow -2t.
 \ee
 Correspondingly we have the identification between the quantum
 numbers:
 \bea\label{identn}
 k&=&-\tilde{\o}_n, \\
 \o&=&2\tilde{k}_n.
 \eea
With this identification, the relation (\ref{squasi}) can be
rewritten as
 \be
 \tilde{\o}_R=-\tilde{k}_n-i2\pi T_Rl(n+h_R)
 \ee
This is reminiscent of the relation (\ref{pole}). The factor $2$
discrepancy with (\ref{pole}) comes from the subtlety in defining
the temperature. Actually, in some literatures, it is $2\pi$
rather than $4\pi$ appeared in (\ref{pole}).

Next, let us consider the quasi-normal modes of the massive vector
field in
 the null warped black hole background.
  One can work with the following first order equations
  \be
   \epsilon_{\lambda}^{\
   \alpha\beta}\partial_{\alpha}A_{\beta}=-mA_{\lambda}.\label{vector}
  \ee
 % The two equations are equivalent only for the Ads$_3$ background.
 %However, we still use (\ref{vector}) for explicit to find some solutions but
 % not all.
 On the null warped black hole background, they become
  \bqa
   \frac{\textrm{d}\phi_t}{\textrm{d}r}&=&\left(-\frac{\omega
   k}{2mlr^2}-\frac{ml}{2r}\right)\phi_t-\frac{\omega^2}{2mlr^2}\phi_{\theta},\\
   \frac{\textrm{d}\phi_{\theta}}{\textrm{d}r}&=&\left(\frac{ml(r^2+r+\alpha^2)}{2r^2}
   +\frac{k^2}{2mlr^2}\right)\phi_t+\left(\frac{ml}{2r}+\frac{\omega^2}{2mlr^2}\right)\phi_\theta\label{phit},\\
   \phi_r&=&\frac{-1}{2mlr^2}(ik\phi_ t+i\omega\phi_\theta),\label{phir}
  \eqa
with the ansatz $A_\mu=e^{-i\omega t+ik\theta}\phi_\mu(r)$. From
the above equations, we obtain a second order differential
equation for $\phi_t$, \be
  \frac{\textrm{d}^2\phi_t}{\textrm{d}
  x^2}+\frac{1}{4}(\frac{\omega^2-m^2l^2+2ml}{x^2}+\frac{\omega^2-2\omega
   k}{x}+\omega^2\alpha^2)\phi_t=0,
\ee
   where $x=\frac{1}{r}$.
It can be solved in terms of Kummer function analogous to the
scalar field case \be
   \phi_t=e^{-\frac{z}{2}}z^{\frac{1}{2}\pm\tilde m_v}F(\frac{1}{2}\pm\tilde m_v-\k,1\pm2\tilde m_v,
   z),
 \ee
 where $z=-i\omegaup\alpha x,\ \tilde
 m_v=\frac{1}{2}\sqrt{(ml-1)^2-\omegaup^2}$ and $
 \k=\frac{i}{4\alpha}(\omegaup-2k)$. Note that $\phi_\theta,\phi_r$
 can be solved straightforwardly by using (\ref{phit}) (\ref{phir}).
  Considering the boundary
 condition for the quasi-normal modes, we also have the similar relation as in the scalar case.
 One also has to make a combination of the two solutions with only ingoing modes at the horizon.
 The coefficients in the combination are
 \be
   C_1=-\frac{\G(1-2\tilde m_v)}{\G(\frac{1}{2}-\tilde m_v-\k)}C,
    \ \ C_2=\frac{\G(1+2\tilde m_v)}{\G(\frac{1}{2}+\tilde
    m_v-\k)}C,
 \ee
 with $C$ being a constant.
 And the flux has a leading term proportional to $C_1^*C_2-c.c.$
 and the sub-leading term $C_2^*C_2r^{-1+2\tilde m_v}$.
 So from the condition that the flux should be finite at asymptotic infinity, we have
% the similar relations \be\label{vquasi}
%   \frac{1}{2}+\tilde m_v-\k=-n,\ \  \textrm{or}\ \ \ \ \frac{1}{2}-\tilde m_v-\k=-n \label{vquasi}
% \ee
%where we have to choose $\tilde
%m_v=\frac{1}{2}\sqrt{(ml-1)^2-\omegaup^2}$.
%if $Re\tilde m_s<\frac{1}{2}$. And if $Re\tilde m_s>\frac{1}{2}$,
%there is only one set of quasi normal modes as follow
\be\label{vquasi}
  \frac{1}{2}+\tilde m_v-\k=-n.
 \ee
 The conformal dimension of the
massive vector field could be obtained easily:
 \be
 h_R^v=\frac{1}{2}+ \frac{1}{2}\sqrt{(ml-1)^2-4\tilde{k}_n^2}.
 \ee
With this and the identification (\ref{identn}), we have the
relation
 \be
 \tilde{\o}_R^v=-\tilde{k}_n-i2\pi T_Rl(n+h_R^v),
 \ee
 from (\ref{vquasi}).

Lastly, we turn to the study of the quasi-normal modes of the
fermionic perturbations. The analysis of the quasi-normal modes
for the fermionic fields in the null warped black hole background
is analogous to the stretched case. We first choose the vielbein:
 \bqa
 e^0&=&rl\textrm{d} t+\frac{1}{2}l(1-\alpha^2-r-r^2)\textrm{d}
 \theta,\nn\\
 e^1&=&\frac{l}{2r}\textrm{d} r,\nn \\
 e^2&=&-rl\textrm{d} t+\frac{1}{2}l(1+\alpha^2+r+r^2)\textrm{d}
 \theta\nn
 \eqa
and then calculate the spin connection. The none-zero components
of the spin connection are
 \bqa
  &&\omega_0^{01}=-\omega_0^{10}=(1+\alpha^2-r^2)/l,\ \
  \omega_0^{12}=-\omega_0^{21}=(\alpha^2-r^2)/l,\ \
  \omega_1^{02}=-\omega_1^{20}=1/l,\nn\\
  &&\omega_2^{01}=-\omega_2^{10}=(\alpha^2-r^2)/l,\ \ \ \
  \omega_2^{12}=-\omega_2^{21}=(-1+\alpha^2-r^2)/l,\nn
 \eqa
 where $\omega_a^{bc}=e_a^{\mu}\omega_{\mu}^{bc}$.
 %\bqa
 % -2r \partial_r\psi_{1}=(1+\frac{1+\alpha^2+r+r^2}{2r}i\omega-i
  %k)\psi_1+(m+\frac{1}{2}+\frac{-1+\alpha^2+r+r^2}{2r}i\omega-i
 % k)\psi_2\\
 % -2r \partial_r\psi_{2}=(m+\frac{1}{2}-\frac{-1+\alpha^2+r+r^2}{2r}i\omega+i
 % k)\psi_1+(1-\frac{1+\alpha^2+r+r^2}{2r}i\omega+i
  %k)\psi_2
 %\eqa

Taking the ansatz $\Psi_i=e^{-i\omega t+ik\theta}\psi_i(r)$ and
making the redefinition of the fields as
 \be
   \psi_1=\psi_1'-\psi_2',\ \ \ \ \psi_2=\psi_1'+\psi_2',
 \ee
we rewrite the Dirac equations as
 \bqa
    \frac{\textrm{d}\psi_1'}{\textrm{d}r}&=&-\frac{2ml+3}{4r}\psi_1'+\frac{i\omega}{2r^2}\psi_2',\label{psi2}\\
     \frac{\textrm{d}\psi_2'}{\textrm{d}r}&=&\frac{i\omega(r^2+r+\alpha^2)-2i k
     r}{2r^2}\psi_1'+\frac{2ml-1}{4r}\psi_2'.
 \eqa
As before we change the variable to $x=\frac{1}{r}$ and redefine
the fields as $P=x^{-\frac{1}{2}}\psi_1'$, then we obtain a second
order differential equation,
 \be
 \frac{\textrm{d}^2 P}{\textrm{d}
x^2}+\frac{1}{4}\left\{\frac{1+\omega^2-(ml-\frac{1}{2})^2}{x^2}+\frac{\omega^2-2\omega
k}{x}+\omega^2\alpha^2\right\}P=0, \ee which can be solved in
terms of Kummer functions
 \be
  \psi_1'=x^{\frac{1}{2}}e^{-\frac{z}{2}}z^{\frac{1}{2}\pm\tilde m_f}
  F(\frac{1}{2}\pm\tilde m_f-\k,1\pm2\tilde m_f, z)
\ee where $\tilde
m_f=\frac{1}{2}\sqrt{(ml-\frac{1}{2})^2-\omega^2},\
\k=\frac{i}{4\alpha}(\omega-2k)$.

 The solution of $\psi_2'$ can be
obtained by using (\ref{psi2}). Similarly considering the boundary
condition for the quasi-normal modes, we have
 \be\label{fquasi}
   \frac{1}{2}+\tilde m_f-\k=-n.
 \ee
%if $Re\tilde m_f>\frac{1}{2}$ and
%\be \frac{1}{2}+\tilde m_f-\k=-n,\
%\ \textrm{or}\ \ \ \frac{1}{2}+\tilde m_f-\k=-n
% \ee
% if $Re\tilde m_f<\frac{1}{2}$

The conformal dimension of the fermion operator is
 \be
 h_R^f=\frac{1}{2}+
 \frac{1}{2}\sqrt{(ml-\frac{1}{2})^2-4\tilde{k}_n^2}.
 \ee
 This relation and the identification (\ref{identn}) give us
\be
 \tilde{\o}^f_R=-\tilde{k}_n-i2\pi T_Rl(n+h^f_R),
 \ee
 from (\ref{fquasi}).

In short, the quasi-normal modes for various perturbations,
including the massive scalar, vector and spin $1/2$ fermion, of
the null warped $AdS_3$ black hole could all be written in a
concise form \be
 \tilde{\o}_R=-\tilde{k}_n-i2\pi T_Rl(n+h_R).
 \ee
This relation is quite similar to the prediction (\ref{pole}) of
warped AdS/CFT correspondence, up to a factor $2$. We take it as
strong evidence to support the conjectured correspondence.

\section{Conformal dimensions}

In this section, we try to compute the conformal dimensions of the
dual operators corresponding to various perturbations around the
spacelike warped and null warped backgrounds. Instead of analyzing
the asymptotic behavior of the solution  of the equation of
motions of the perturbations, we take a slightly more algebraic
way. For both the spacelike stretched $AdS_3$ and the null warped
$AdS_3$, they have the isometry group $SL(2,R)\times U(1)$. The
perturbations should respect the isometry group. The highest
conformal weight mode created by the bulk perturbations must obey
the algebraic equation $L_1\phi=0$. Its $L_0$ eigenvalue can be
taken as
 the conformal dimensions of the
primary operators in the boundary CFT.
 It turns out that the conformal dimensions determined in this way
 are in perfect agreement with the ones obtained before.  For simplicity, we just focus on the
 scalar and the vector
perturbations.

\subsection{Spacelike stretched case}

 Let us consider a massive scalar $\Phi$ of mass $m$ in the warped
spacelike $AdS_3$ spacetime. From warped AdS/CFT correspondence,
such a scalar field have a counterpart boundary operator in dual
conformal field theory. We work in the following form of the
spacelike stretched $AdS_3$:
 \be
 ds^2=\frac{l^2}{v^2+3}\left(-\cosh^2\s d\t^2+d\s^2+\frac{4v^2}{v^2+3}(du+\sinh\s
 d\t)^2\right),
 \ee
 which is the same as (\ref{spacelikemetric}) by the coordinate
 transformation $r=\sinh\s, x=u$.

Such a background has the $U(1)_L\times SL(2,R)_R$ isometries. The
 $U(1)_L$ isometry is generated by
 \be
 L_0=i\p_u,
 \ee
 and the $SU(2)_R$ isometry is generated by $\bar{L}_0,\bar{L}_1$
 and $\bar{L}_{-1}$ satisfying
 \be
 [\bar{L}_0, \bar{L}_{\pm 1}]=\mp\bar{L}_{\pm 1},
 \hspace{3ex}
 [\bar{L}_1, \bar{L}_{-1}] = 2\bar{L}_0,
 \ee
 where
 \bea
 \bar{L}_0&=&i\p_\t, \\
 \bar{L}_1&=&-e^{i\t}(\p_\s+i\tanh\s\p_\t+i\frac{1}{\cosh\s}\p_u),\\
 \bar{L}_{-1}&=&e^{-i\t}(\p_\s-i\tanh\s\p_\t-i\frac{1}{\cosh\s}\p_u).
 \eea

The scalar equation of motion now takes the form
 \bea
& & \frac{1}{\cosh\s}\p_\s(\cosh\s\p_\s)\Phi-\frac{1}{\cosh^2\s}\p^2_\t\Phi+\frac{2\sinh\s}{\cosh^2\s}\p_\t\p_u\Phi \nn\\
& &
-\frac{\sinh^2\s}{\cosh^2\s}\p^2_u\Phi+\large(\frac{v^2+3}{4v^2}\large)\p^2_u\Phi-\frac{m^2l^2}{v^2+3}\Phi=0.
\eea The above equation could be rewritten as
 \be
 \left\{-[\frac{1}{2}(\bar{L}_1\bar{L}_{-1}+\bar{L}_{-1}\bar{L}_1)-\bar{L}^2_0]
 +\frac{3(v^2-1)}{4v^2}L^2_0-\frac{m^2l^2}{v^2+3}\right\}\Phi=0.
 \ee
One may make the following ansatz \be
\Phi=e^{-i\tilde{\o}_R\t+i\tilde{k}u}\phi. \ee The corresponding
highest weight mode should satisfy
 \be\label{Phi}
 \bar{L}_1\Phi=0,\hspace{4ex}\bar{L}_0\Phi=h_R\Phi.
 \ee
 This helps us to fix the conformal dimension\footnote{The
 conformal weights of the scalar fields have been discussed in
 \cite{Anninos2009} in the similar way independently. We would like to thank the
 anonymous referee for pointing this out.}
 \be
 h_R=\tilde \o_R=\frac{1}{2}\pm \sqrt{\frac{1}{4}+s},
 \ee
 where $s$ is defined to be (\ref{s}). This is the same as
 (\ref{sweight}), where we pick $+$.

 For the highest weight state, we can even solve the equation
 (\ref{Phi}) and get
 \be
 \Phi=e^{-ih_R\t+i\tilde{k}u}e^{\tilde{k}\tan^{-1}\sinh\s}(\cosh\s)^{-h_R}.
 \ee

In order to study the conformal weight for the vector fields in
the  space-like warped
 AdS$_3$,
   it is more convenient to use the Poincare coordinates \cite{Andy08}
 \be
   ds^2=\frac{l^2}{v^2+3}(-x^2\textrm{d}t^2+
   \frac{\textrm{d}x^2}{x^2}+\frac{4v^2}{v^2+3}(\textrm{d}\th+x\textrm{d}t)^2).
 \ee
  The Killing vectors of the space-like warped AdS$_3$ are given by
  \bqa
    V_{-1}=i(-\frac{1}{x^2}-t^2)\p_t+2itx\p_x+\frac{2i}{x}\p_\th,\\
    V_0=t\p_t-x\p_x,\ \ V_1=i\p_t,\ \ V=i\p_{\th},
  \eqa
   which satisfy the commutation relations:
 \bea
 [V_0,V_{\pm 1}]=\mp V_{\pm 1}, \hspace{5ex} [V_1,V_{-1}]=2V_0.
 \eea
For the highest weight state created by the vector field, it
 satisfies
 \bqa
   \mathcal {L}_{V_1}A_{\mu}=0,\hspace{4ex} \mathcal {L}_{V_0}A_{\mu}=h_R^vA_{\mu},\hspace{4ex}
    \mathcal {L}_VA_{\mu}=-\tilde kA_{\mu},
 \eqa
  where  $\mathcal {L}$ denotes Lie derivative. Then the solution is
  \be
    A_t=C_1x^{1-h^v_R}e^{i\tilde{k}\th},\hspace{4ex}
    A_x=C_2x^{-1-h^v_R}e^{i\tilde{k}\th},
    \hspace{4ex}A_{\th}=C_3x^{-h^v_R}e^{i\tilde{k}\th}.
  \ee
Using the equation of motion for the vector fields, we obtain
 \bqa
   C_2=-\frac{i\tilde{k}(v^2+3)}{2m l v}C_1,\hspace{4ex}
   C_3=\frac{2m l v(h-1)-\tilde{k}^2{(v^2+3)}}{m l(2hv-m l)}C_1
   \eqa
   and
   \be
   h_R^v=\frac{1}{2}+
 \sqrt{\frac{1}{4}+\frac{m^2l^2-2m l
 v}{v^2+3}-\frac{3(v^2-1)\tilde{k}^2}{4v^2}}.
 \ee
 This is slightly different from (\ref{vweight}) by the sign before
 $m$, but is exactly the same as (\ref{hRv}). The difference comes from
 that through the coordinate
 transformation, the helicity also exchange, which induce
 $m\rightarrow -m$.

It is interesting to compare the above results with the ones in
usual $AdS_3/CFT_2$ correspondence. In latter case, one has
 \be\label{weight}
 h_R+h_L=\Delta, \hspace{3ex} h_R-h_L=\pm s,
 \ee
 where $s$ is the spin of the field,
 \be\label{dim1}
 \Delta=1\pm\sqrt{1+m^2l^2},
 \ee
 for the scalar fields, and
 \be\label{dim2}
 \Delta=1+|m|l,
 \ee
 for the vector fields. For warped $AdS_3/CFT_2$ correspondence, the
 relations (\ref{weight}) still make sense, even though we cannot determine
 $h_L$ directly. Another property shared by both cases is that in the vector and
 the fermionic case, when the helicity changes, $m$ changes sign and accordingly the
 expression of the conformal dimension need to be changed slightly.
 However, the relations
 (\ref{dim1},\ref{dim2}) have to be modified greatly in the warped AdS case. One
 modification is on the scale, from $l$ to $2l/\sqrt{v^2+3}$.
 Another modification is more significant: in the warped case, another
 quantum number from $U(1)_R$ appears in the conformal dimensions.
 This is not only true for the spacelike stretched case but also true
 for the null warped case.
As a
 consistent check, when $v=1$, our result reduce to
 (\ref{dim1},\ref{dim2}).

 One interesting feature in the conformal weights of operators in
 dual CFT is that they depend on the $U(1)$ quantum number
 $\tilde{k}$. The presence of this quantum number means that even
 though the mass-square of the scalar field satisfies the
 Breitenlohner-Freedman bound for three-dimensional AdS spacetime,
 the perturbation could still be unstable. As a result,
 superradiance may happen in the spacelike stretched AdS$_3$
 spacetime\cite{Anninos2009}, just like in Kerr black hole\cite{Bredberg:2009pv}.
 Similar phenomenon happens in the null warped AdS$_3$
 spacetime as well.

\subsection{Null warped case}

 For the null warped $AdS_3$
spacetime (\ref{nullmetric}), it has isometry group $SL(2,R)_R
\times U(1)_{null}$. The $U(1)_{null}$ is generated by
 \be
 N=\p_+,
 \ee
 and $SL(2,R)_R$ is generated by
 \bea
 N_1&=&\p_-,\nn\\
 N_0&=&x^-\p_-+\frac{u}{2}\p_u,\nn\\
 N_{-1}&=&(x^-)^2\p_--u^2\p_++x^-u\p_u, \nn
 \eea
 which satisfy the commutation relations:
 \bea
 [N_0,N_{\pm 1}]=\mp N_{\pm 1}, \hspace{5ex} [N_1,N_{-1}]=2N_0.
 \eea

 The equation of motion of the massive scalar $\Phi$ in the null warped
 $AdS_3$ spacetime is of the form
 \be
 (u^2\p_u^2-u\p_u-4\p^2_++4u^2\p_+\p_--m^2l^2)\Phi=0,
 \ee
which could be rewritten as
 \be
 [\frac{1}{2}(N_1N_{-1}+N_{-1}N_1)-N^2_0+N^2+\frac{m^2l^2}{4}]\Phi=0.
 \ee
The highest conformal weight state should satisfy
 \be
 N_1 \Phi=0,\hspace{4ex} N_0\Phi=h_R^s\Phi,\hspace{4ex} N\Phi=i\tilde
 k_n\Phi.
 \ee
This just gives the constraint $\tilde{\o}=0$ and the conformal
weight $h_R^s=\frac{1}{2}\pm
 \frac{1}{2}\sqrt{1+m^2l^2-4\tilde{k}_n^{2}}$. The solution
 of the highest weight state is
 \be
   \Phi=u^{2h_R}e^{i\tilde k_nx^+}
 \ee
 The consistent boundary condition require $h_R^s>0$.

 For the conformal weight of the vector field, the highest weight state
 satisfy
 \bqa
   \mathcal {L}_{N_1}A_{\mu}=0,\hspace{4ex} \mathcal {L}_{N_0}A_{\mu}=h_R^vA_{\mu},\hspace{4ex}
    \mathcal {L}_NA_{\mu}=i\tilde k_nA_v.\label{nullvector}
 \eqa
 This is consistent with
  the equation of motion.  For any Killing vector $\xiup$, we can choose a
  coordinate $y$ satisfying $\xiup=\p_y$. In this special
  coordinates, the background metric is independent of $y$. The
  equation of motion for the vector fields can be seen as a linear
  operator acting on the vector fields. Since the operator  only
  depends on the metric, the operator commutes with the Lie
  derivative of the Killing vector. So the equations for the highest weight solution
  are   consistent with the equations of the vector fields.
  The solution of (\ref{nullvector} is
  \be
    A_+=C_1u^{2h_R^v}e^{i\tilde
    k_nx^+},\hspace{4ex}
    A_-=C_3u^{2h_R^v}e^{i\tilde k_nx^+},
    \hspace{4ex}A_{u}=C_2u^{2h_R^v-1}e^{i\tilde k_nx^+}.
  \ee
Using the equation of motion for the vector fields, we obtain \bqa
C_1=\frac{-2\tilde k_n^{2}}{m l(h_R^v-m l)}C_2,\hspace{4ex}
C_3=\frac{-2i\tilde k_n}{m l}C_2,\eqa
  and also find the conformal dimension
  \be\label{nullv}
 h_R^v=\frac{1}{2}+
 \frac{1}{2}\sqrt{(ml-1)^2-4\tilde{k}_n^{2}}.
 \ee

Comparing with the conformal dimensions in usual AdS/CFT
correspondence, we see that the only difference is the appearance
of $\tilde{k}$ terms in (\ref{nullv}).

\section{Conclusions and discussion}

In this paper, we calculated the quasi-normal modes of various
perturbations, including the massive scalar, vector and spin
one-half fermionic perturbations, of the spacelike stretched and
the null warped $AdS_3$ black holes. For the spacelike stretched
black hole, all kinds of the quasi-normal modes could be rewritten
in terms of the quantum numbers $\tilde{\o}$ and $\tilde{k}$ in a
simple way:
 \bea\label{quasistr}
\tilde{\o}_R&=&\frac{1}{v^2+3}(-4\pi T_Ll\tilde{k}-(i 4\pi
 T_Rl)(n_1+h_R)),\hspace{3ex}\mbox{or}\nn\\
 \tilde{k}&=&-i(n_2+h_L),
 \eea
 where $n_1,n_2$ are non-negative integers.
 Similarly, for the null warped black hole, the quasi-normal
modes are of the form
 \be\label{quasinull}
 \tilde{\o}_R=-\tilde{k}_n-i2\pi T_Rl(n+h_R),
 \ee
 with $n$ being non-negative integer.
 The above relations are reminiscent of the relations (\ref{pole}) on
 the poles of the retarded Green's function.
 Since the conjectured correspondence is between the spacelike
 stretched  (null) warped $AdS_3$ and its holographically dual 2D CFT,
 one needs to use the
 quantum numbers appeared in these spacetimes rather than the ones
 in the black holes to set up the dictionary. This is why we rewrite the quasi-normal modes in
 terms of the quantum numbers $\tilde{\o}$ and $\tilde{k}$ of the spacelike (null) warped $AdS_3$.
 And actually it is in terms of these quantum numbers that make the warped AdS/CFT
 correspondence manifest. This is the key point to set up the
 dictionary. The phenomena happening here is extraordinary. The asymptotic
geometries of the warped black holes could be locally transformed
to the ones of the warped spacetimes. The coordinate
transformations induce the identifications of two sets of quantum
numbers. Taken this subtlety into account, the quasi-normal modes
of the warped black holes could be reorganized into
(\ref{quasistr},\ref{quasinull}), which are well consistent with
the CFT prediction (\ref{pole}) and so support the
 conjectures.

The relations (\ref{quasistr},\ref{quasinull}) were obtained after
 the identifications (\ref{ident},\ref{identn}) being taken into account.
 However, the
identifications come from the local transformations, rather than
the global ones. Especially, considered the fact that the global
warped AdS$_3$ spacetime is not included in the black hole's phase
space, the above identifications deserve further investigations
and clarifications. We wish we can return to this issue in the
future.

One interesting point is that it seems that we have only one set
of the quasi-normal modes. For the null warped case, this seems to
be natural since the dual CFT has only right temperature. For the
spacelike stretched case, this sounds strange. However, recall
that the isometry group of the spacelike stretched $AdS_3$ is just
$SL(2,R)_R\times U(1)_L$. For the highest weight operators
corresponding to the massive scalar, one can define its
right-moving conformal weight from $SL(2,R)_R$, but can only
define $\tilde{k}$ from $U(1)_L$. This in fact is in consistence
with (\ref{quasistr}).

 Another interesting point in our result is that the above two
 relations (\ref{quasistr},\ref{quasinull}) are still a little
 different from the predicted poles (\ref{pole}), up to a scale
 factor. This could be due to the ambiguity in determining the
 temperature, originated from coordinate transformation. This possibility
 has been shown in \cite{ChenXu09} for comparison with the BTZ black hole. It would
 be interesting to pin down this issue.

In this paper, we proposed a conjecture that the quantum
topological massive gravity asymptotic to the null warped AdS$_3$
is holographically dual to 2D CFT. Our study on the quasi-normal
modes of the null warped black hole support this conjecture.
However it would be essential to put this conjecture on a more
solid ground. One interesting issue is that the holographic
anomaly suggests that there should be not only right sector but
also left one as well. This could not be seen from the study of
the black hole thermodynamics and the quasi-normal modes. To
understand this issue better, it would be important to investigate
the asymptotical boundary conditions on the gravitational
perturbations and check if the central charges could be derived
from the symmetry algebra.

% Another interesting issue is on conformal dimensions. In
 %appendix, we calculate the conformal dimension of scalar operator
 %from Virasoro algebra for spacelike stretched $AdS_3$. It is the same as the one we obtain from solving the scalar
 %equation at asymptotic region. It would be interesting to decide the
 %conformal dimensions of vector and fermionic operators in the
 %same algebraic way. On the other hand, for the null warped
 %$AdS_3$, one cannot obtain the conformal weight in the algebraic
 %way, even for scalar operators.

 It would be worth looking for other evidence to support the
 warped AdS/CFT correspondence. One possibility is to compare the
 absorption cross sections. This has been explored in the context of
  Kerr/CFT correspondence\cite{Bredberg:2009pv}. For the warped
  $AdS_3$ black hole, since the equation of motion of
the  perturbations are exactly solvable, we expect that the same
  analysis would be feasible\cite{ChenNingXu}.

In this paper, we discussed the quasi-normal modes of the scalar,
vector and fermionic perturbations. It would be interesting to
consider the gravitational perturbations. In this case, the
equations of motions is a third order differential equation, so is
more difficult to solve. Nevertheless, It was shown in
\cite{Anninos:2009zi} that after fixing the gauge completely, the
equations of motion in the warped AdS$_3$ could be simplified. We
expect the same simplification may happen in the warped black hole
case.

\section*{Acknowledgments}

We would like to thank R.G. Cai, J.X. Lu and B. Wang for valuable
discussions and comments.  BC would like to thank KIAS for
hospitality during his visit.
 The work was partially supported by NSFC Grant
No.10535060,10775002,10975005 and NKBRPC (No. 2006CB805905).

\ed

\end{document}